\documentstyle[12pt]{article}
\begin{document}
\vspace*{2cm}
\begin{center}
 \Huge\bf
A new stationary cylindrically symmetric solution \\ of the Einstein's 
equations  \\ admiting Time machine
\vspace*{0.25in}

\large

Elena V. Palesheva
\vspace*{0.15in}

\normalsize

Department of Mathematics, Omsk State University \\
644077 Omsk-77 RUSSIA
\\
\vspace*{0.5cm}
E-mail: m82palesheva@math.omsu.omskreg.ru  \\
\vspace*{0.5cm}
July 28, 2001\\
\vspace{.5in}
ABSTRACT
\end{center}

In this article a new stationary solution of the Einstein's
equations with cosmological constant and Time machine   is given. The
garavitational field is created by ideal liquid with three massless scalar
fields or by ideal liquid with electric-magnetic field and massless scalar
field.

\newpage

\setcounter{page}{1}


\section{Introduction}
In this article a new stationary solution of the Einstein's
equations with cosmological constant and Time machine   is given.
There are two interpretations of stress-energy tensor for this spacetime. In
the first case cosmological constant can be arbitrary and 
gravitational field is  created
by ideal liquid with three massless scalar fields. In second case the 
cosmological
constant is non-negative and matter which creates this gravitational field 
is an ideal liquid with electric-magnetic field and massless scalar field.
In 1937 the first similar solution was found by W.J.~van Stockum
\cite{1}. But problem of Time machihe is discoursed from 1949, after that
cosmological model admiting smooth closed timelike curves was found by
the  famous logic
K.~G\"odel \cite{2}. He was the first who interpreted similar curves as Time
machine.

\section{Metric and stress-energy tensor}
Let us look the metric
$$
 ds^2=\frac{{dx^{\scriptscriptstyle 0}}^2}{2\Omega}+2
(x^{\scriptscriptstyle 2}dx^
 {\scriptscriptstyle 1}-x^{\scriptscriptstyle 1}dx^{\scriptscriptstyle 2})
dx^
 {\scriptscriptstyle 0}+\Omega (2{x^{\scriptscriptstyle 2}}^2-1)
{dx^{\scriptscriptstyle 1}}^2+
 \Omega (2{x^{\scriptscriptstyle 1}}^2-1){dx^{\scriptscriptstyle 2}}^2-
$$
$$
 -4\Omega x^
 {\scriptscriptstyle 1}x^{\scriptscriptstyle 2}dx^{\scriptscriptstyle 1}
dx^{\scriptscriptstyle
 2}-{dx^{\scriptscriptstyle 3}}^2,\eqno (1)
$$
where $\Omega=const>0$. The nonzero components of the Christoffel's symbols are
$$
\Gamma^0_{01}=2x^{\scriptscriptstyle 1},\,\Gamma^0_{02}=
2x^{\scriptscriptstyle 2}
,\,\Gamma^0_{11}=8x^{\scriptscriptstyle 1}x^{\scriptscriptstyle 2}
\Omega,
\,\Gamma^0_{12}=4\Omega ({x^{\scriptscriptstyle 2}}^2-
{x^{\scriptscriptstyle 1}}^2),
\,\Gamma^0_{22}=-8x^{\scriptscriptstyle 1}x^{\scriptscriptstyle 2}\Omega,
$$
$$
\Gamma^1_{02}=-\frac{1}{\Omega},\,\Gamma^1_{12}=-2x^{\scriptscriptstyle 2},\,
\Gamma^1_{22}=4x^{\scriptscriptstyle 1},\,\Gamma^2_{01}=\frac{1}{\Omega}
,\,\Gamma^2_{11}=4x^{\scriptscriptstyle 2},\,\Gamma^2_{12}=
-2x^{\scriptscriptstyle 1},
$$
and nonzero components of the Ricci tensor are
$$
R_{00}=\frac{2}{{\Omega}^{\lefteqn{\scriptstyle 2}}},\,R_{01}=\frac{4x^{
\scriptscriptstyle 2}}{\Omega},\,R_{02}=
\frac{-4x^{\scriptscriptstyle 1}}{\Omega}
,\,R_{11}=4+8{x^{\scriptscriptstyle 2}}^2,\,R_{12}=-8x^{
\scriptscriptstyle 1}x^{\scriptscriptstyle 2},\,R_{22}=
4+8{x^{\scriptscriptstyle 1}}^
2.
$$
Also was calculated scalar curvature $R=-4/\Omega$.

By using the Einstein's equations\footnote {The Greek indexes are 1,2,3, and
Latin indexes are 0,1,2,3.}
$$
R_{ik}-\frac{1}{2} g_{ik}R=\kappa T_{ik}+\Lambda g_{ik},
$$
we find the following tensor:
$$
\kappa T_{ik}+\Lambda g_{ik}=\left[\begin{array}{cccr}
{\displaystyle \frac{3}{{\mathstrut\Omega}^{\lefteqn{\scriptstyle2}}}}&
{\displaystyle \frac{6x^{\scriptscriptstyle 2}}{\Omega}}&
{\displaystyle -\frac{6x^{\scriptscriptstyle 1}}{\Omega}}&0\\ 
{\displaystyle \frac{{\mathstrut 6x}
^{\scriptscriptstyle 2}}{\mathstrut 
\Omega}}&12{x^{\scriptscriptstyle 2}}^2+2&-12x^{
\scriptscriptstyle 1}x^{\scriptscriptstyle 2}&0\\ 
{\displaystyle -\frac{{\mathstrut 6x}^{
\scriptscriptstyle 1}}{\Omega}}&-
12x^{\scriptscriptstyle 1}x^{\scriptscriptstyle 2}&12{x^{
\scriptscriptstyle 1}}^2+2&0\\ 0&0&0&{\displaystyle -\frac{2}{\Omega}}
\end{array}\right].\eqno (2)
$$
Note that if we introduce the cylindrical coordinats
$$
\left\{\begin{array}{l}x^{\scriptscriptstyle 0}=
x^{\scriptscriptstyle 0}\\ x^{
\scriptscriptstyle 1}=r\cos \varphi\\ x^{\scriptscriptstyle 2}=
r\sin \varphi\\ x^{\bigskip
\scriptscriptstyle 3}=x^{\scriptscriptstyle 3}\end{array}\right.
$$
then metric is transformed to the expression
$$
ds^2=\frac{1}{2\Omega}d{x^{\scriptscriptstyle
0}}^2-2r^2dx^{\scriptscriptstyle 0}d\varphi- \Omega dr^2+\Omega
r^2\left(2r^2-1\right)d{\varphi}^2-d{x^{\scriptscriptstyle 3}}^2.
$$

\section{The physical interpretations of the stress-energy tensor}

\subsection{Interpretation 1}

In this section we show that geometry of spacetime with metric (1) can be
created by ideal liquid, for which we must take
$$
\left\{\begin{array}{l}
T_{{\mathstrut ik}}^{(i.liquid)}=(c^2\rho +p)u_iu_k-p\,g_{ik}\\
g^{{\mathstrut ik}}u_iu_k=1\end{array}\right. ,\eqno (3)
$$
and for scalar fields
$$
T_{ik}^{scalar}=\frac{\partial\varphi}{\partial x^{\scriptscriptstyle i}}
\frac{\partial\varphi}{\partial
x^{\scriptscriptstyle k}}+\frac{1}{2}g_{ik}(m^2{\varphi}^2-
g^{mn}\frac{\partial\varphi}{\partial
x^{\scriptscriptstyle m}}\frac{\partial\varphi}{\partial x^
{\scriptscriptstyle n}}),\eqno (4)
$$
$$
-\frac{1}{\sqrt{-g}}\,\frac{\partial}{\partial x^{\scriptscriptstyle i}}
(\sqrt{-g}\,g^{ik}\frac
{\partial\varphi}{\partial x^{\scriptscriptstyle k}})-m^2\varphi=0.\eqno (5)
$$
Here (5) is the  Klein-Fock's equation.

Let there are three real massless scalar fields and ideal liquid. According
to these equations  gravitational field will be determinated by
equality
$$
\kappa T_{ik}+\Lambda g_{ik}=\kappa (c^2\rho +p)u_iu_k+\kappa\frac{\partial
\varphi}{\partial x^{\scriptscriptstyle i}}\frac{\partial\varphi}{\partial
x^{\scriptscriptstyle k}}+\kappa\frac{\partial\psi}{\partial
x^{\scriptscriptstyle i}}\frac{\partial\psi}{\partial x^{\scriptscriptstyle k}}
+\kappa\frac{\partial\theta}{\partial x^{\scriptscriptstyle i}}\frac{\partial
\theta}{\partial x^{\scriptscriptstyle k}}+g_{ik}\{\Lambda -\kappa p-
$$
$$
-\frac{
\kappa}{2}g^{mn}(\frac{\partial\varphi}{\partial x^{\scriptscriptstyle m}}
\frac{\partial\varphi}{\partial x^{\scriptscriptstyle n}}+\frac{\partial\psi}
{\partial x^{\scriptscriptstyle m}}\frac{\partial\psi}{\partial x^{
\scriptscriptstyle n}}+\frac{\partial\theta}{\partial x^{\scriptscriptstyle m}}
\frac{\partial\theta}{\partial x^{\scriptscriptstyle n}})\}.\eqno (6)
$$
Now we assume that $\varphi=\varphi(x^{\scriptscriptstyle 1})$, $ \psi=
\psi(x^{
\scriptscriptstyle 2})$ and $\theta=\theta(x^{\scriptscriptstyle 3})$, or 
that only $ \partial\varphi/\partial x^{\scriptscriptstyle 1}, \partial\psi/
\partial x^{\scriptscriptstyle 2}$ and $ \partial\theta/\partial x^{
\scriptscriptstyle 3}$ are not equial to zero. The vector
$$
u_i=(\pm {\displaystyle \frac{1}{\sqrt{2\Omega}}},\pm \sqrt{2\Omega}x^{
\scriptscriptstyle 2},\mp\sqrt{2\Omega}x^{\scriptscriptstyle 1},0)\eqno (7)
$$
satisfies to the restriction (3) for 4-velocity. By using these conjectures 
and (5), and substituting (2) into (6) we obtain
$$
\begin{array}{l}{\displaystyle \frac{3}{{\mathstrut\Omega}^{\lefteqn{
\scriptstyle 2}}}=\frac{\kappa (c^2\rho +p)}{2\Omega}+\frac{1}{2\Omega}\left(
\Lambda -\kappa p+\frac{\kappa}{2\Omega}{\left(\frac{\partial\varphi}{\partial
x^{\scriptscriptstyle 1}}\right)}^2+\frac{\kappa}{2\Omega}{\left(\frac{
\partial\psi}{\partial x^{\scriptscriptstyle 2}}\right)}^2+\right. }\\
      \hspace*{1.5cm}{\displaystyle \left. +\frac{\kappa}{2
\Omega}{\left(\frac{\partial\theta}{\partial x^{\scriptscriptstyle 3}}\right)}
^2\right)}\\
{\displaystyle \frac{{\mathstrut}6x^{\scriptscriptstyle 2}}{\mathstrut \Omega}
=\kappa (c^2\rho +p)x^{\scriptscriptstyle 2}+x^{\scriptscriptstyle 2}\left(
\Lambda-\kappa p+\frac{\kappa}{2\Omega}{\left(\frac{\partial\varphi}{\partial
x^{\scriptscriptstyle 1}}\right)}^2+\frac{\kappa}{2\Omega}{\left(\frac{
\partial\psi}{\partial x^{\scriptscriptstyle 2}}\right)}^2+\right. }\\
    \hspace*{1.5cm}{\displaystyle \left. +\frac{\kappa}{2
\Omega}{\left(\frac{\partial\theta}{\partial x^{\scriptscriptstyle 3}}\right)
}^2\right)}\\
{\displaystyle -\frac{{\mathstrut 6x}^{\scriptscriptstyle 1}}{\mathstrut
\Omega}=-\kappa (c^2\rho+p)x^{\scriptscriptstyle 1}-x^{\scriptscriptstyle 1}
\left(\Lambda -\kappa p+\frac{\kappa}{2\Omega}{\left(\frac{\partial\varphi}{
\partial x^{\scriptscriptstyle 1}}\right)}^2+\frac{\kappa}{2\Omega}{\left(
\frac{\partial\psi}{\partial x^{\scriptscriptstyle 2}}\right)}^2+\right. }\\
    \hspace*{1.5cm}{\displaystyle  \left.+\frac{\kappa
}{2\Omega}{\left(\frac{\partial\theta}{\partial x^{\scriptscriptstyle 3}}
\right)}^2\right)}\\
{\displaystyle {\mathstrut -12x^{\scriptscriptstyle 1}x^{\scriptscriptstyle 2
}}=-2\Omega\,x^{\scriptscriptstyle 1}x^{\scriptscriptstyle 2}\kappa 
(c^2\rho +
p)-2\Omega\,x^{\scriptscriptstyle 1}x^{\scriptscriptstyle 2}\left(\Lambda -
\kappa p+\frac{\kappa}{2\Omega}{\left(\frac{\partial\varphi}{\partial x^{
\scriptscriptstyle 1}}\right)}^2+\right.}\\
    \hspace*{1.5cm}{\displaystyle \left.+\frac{\kappa}{2\Omega}
{\left(\frac{\partial
\psi}{\partial x^{\scriptscriptstyle 2}}\right)}^2 +\frac{\kappa}{2\Omega}{
\left(\frac{\partial\theta}{\partial x^{\scriptscriptstyle 3}}
\right)}^2\right)}\\
{\displaystyle {\mathstrut 12{x^{\scriptscriptstyle 2}}^2}+2=2\Omega{x^{
\scriptscriptstyle 2}}^2\kappa (c^2\rho +p)+\kappa{\left(\frac{\partial
\varphi}{\partial x^{\scriptscriptstyle 1}}\right)}^2+\Omega ({2x^{
\scriptscriptstyle 2}}^2-1)\left(\Lambda -\kappa p+\right.}\\
    \hspace*{1.5cm}{\displaystyle \left.+\frac{\kappa}{2\Omega}{
\left(\frac{\partial\varphi}{\partial x^{\scriptscriptstyle 1}}\right)}^2+
\frac{\kappa}{2\Omega}{\left(\frac{\partial\psi}
{\partial x^{\scriptscriptstyle
2}}\right)}^2+\frac{\kappa}{2\Omega}
{\left(\frac{\partial\theta}{\partial x^{
\scriptscriptstyle 3}}\right)}^2\right)}\\
\end{array}
$$
$$
\begin{array}{l}
{\displaystyle {\mathstrut 12{x^{\scriptscriptstyle 1}}^2}+2=2\Omega{x^{
\scriptscriptstyle 1}}^2\kappa (c^2\rho +p)+\kappa{\left(\frac{\partial\psi}{
\partial x^{\scriptscriptstyle 2}}\right)}^2+\Omega ({2x^{\scriptscriptstyle
1}}^2-1)\left(\Lambda -\kappa p+\right.}\\
     \hspace*{1.5cm}{\displaystyle \left.+\frac{\kappa}
{2\Omega}{\left(\frac{\partial
\varphi}{\partial x^{\scriptscriptstyle 1}}\right)}^2+\frac{\kappa}{2\Omega}
{\left(\frac{\partial\psi}{\partial x^{\scriptscriptstyle 2}}
\right)}^2+\frac
{\kappa}{2\Omega}{\left(\frac{\partial\theta}
{\partial x^{\scriptscriptstyle 3
}}\right)}^2\right)}\\
{\displaystyle -\frac{\mathstrut 2}{\mathstrut \Omega}=
\kappa{\left(\frac{
\partial\theta}{\partial x^{\scriptscriptstyle 3}}\right)}^2-
\left(\Lambda -\kappa
p+\frac{\kappa}{2\Omega}{\left(\frac{\partial\varphi}
{\partial x^{\scriptscriptstyle
1}}\right)}^2+\frac{\kappa}{2\Omega}{\left(\frac{\partial\psi}{\partial x^{
\scriptscriptstyle 2}}\right)}^2+\frac{\kappa}{2\Omega}{\left(\frac{\partial
\theta}{\partial x^{\scriptscriptstyle 3}}\right)}^2\right)}\\
{\displaystyle \frac{\partial}{\partial x^{\scriptscriptstyle k}}
\left(g^{1k}\frac{
\partial\varphi}{\partial x^{\scriptscriptstyle 1}}\right)=
\frac{\partial}{\partial
x^{\scriptscriptstyle k}}\left(g^{2k}\frac{\partial\psi}{\partial x^{
\scriptscriptstyle 2}}\right)=
\frac{\partial}{\partial x^{\scriptscriptstyle k}}\left(g^
{3k}\frac{\partial\theta}{\partial x^{\scriptscriptstyle 3}}\right)=0}.
\end{array}
$$
The next formulas are directly checked:
$$
\left\{\begin{array}{l}
\kappa p=\Lambda+{\displaystyle \frac{2}{\mathstrut\Omega}+\frac{\kappa}{2}{
\left(\frac{\partial\theta}{\partial x^{\scriptscriptstyle 3}}\right)}^2}\\
\kappa c^2\rho={\displaystyle \frac{\mathstrut 2}{\mathstrut\Omega}-\Lambda-
\frac{3}{2}\kappa{\left(\frac{\partial\theta}
{\partial x^{\scriptscriptstyle 3}}
\right)}^2}\\
u_i=(\pm {\displaystyle \frac{\mathstrut 1}{\mathstrut\sqrt{2\Omega}}},\pm
\sqrt{2\Omega}x^{\scriptscriptstyle 2},\mp 
\sqrt{2\Omega}x^{\scriptscriptstyle
1},0)\\
{\displaystyle \kappa{\left(\frac{\mathstrut\partial\varphi}
{\mathstrut\partial x^{
\scriptscriptstyle 1}}\right)}^2=
\kappa{\left(\frac{\partial\psi}{\partial x^{
\scriptscriptstyle 2}}\right)}^2=
4+\kappa\Omega{\left(\frac{\partial\theta}{\partial x^{
\scriptscriptstyle 3}}\right)}^2} \\
{\displaystyle \mathstrut\varphi=A_1x^{\scriptscriptstyle 1}+A_2, \psi=
B_1x^{
\scriptscriptstyle 2}+B_2, \theta=C_1x^{\scriptscriptstyle 3}+C_2}\\
A_1,A_2,B_1,B_2,C_1,C_2=const.\end{array}\right.\eqno (8)
$$
So cosmological constant must change in the following domain
$$
{\displaystyle -\frac{2}{\Omega}-
\frac{\kappa}{2}{\left(\frac{\partial\theta}{
\partial x^{\scriptscriptstyle 3}}\right)}^2\leq\Lambda\leq\frac{2}
{\Omega}-\frac{3}
{2}\kappa{\left(\frac{\partial\theta}{\partial x^{\scriptscriptstyle 3}}
\right)}^2}.
$$

\subsection{Interpretation 2}

Above one of the some interpretations of stress-energy tensor was considered. 
Here we shall attempt to demonstrate  another variant of matter. We call
attention to the well-known fact that electric-magnetic stress-energy tensor
$$
T_{ik}^{el.mag.}=\frac{1}{4\pi}\left(\frac{1}{4}F_{lm}F^{lm}g_{ik}-F_{il}F^
{\;l}_k\right).\eqno (9)
$$
Together with (9) we must take the Maxwell's equations
$$
\left\{\begin{array}{l}F_{ik}={\displaystyle \frac{\partial A_k}{\partial x^
{\scriptscriptstyle i}}-\frac{\partial A_i}{\partial x^{\scriptscriptstyle k}
}}\\
{\nabla}_kF^{ik}={\displaystyle -\frac{4\pi}{c}j^{\,i}},\end{array}
\right.\eqno (10)
$$
where $A_k$ is  4-potential of this electric-magnetic field and
${\nabla}_k$ is a covariant derivation.

As early we consider ideal liquid and real massless scalar field.
Moreover we shall use the  
electric-magnetic field. And now we show that considered field's system 
satisfies to (2). We have
$$
\kappa T_{ik}+\Lambda g_{ik}=\kappa(c^2\rho+p)u_iu_k+\kappa\frac{\partial
\varphi}{\partial x^{\scriptscriptstyle i}}\frac{\partial\varphi}{\partial x^
{\scriptscriptstyle k}}-\frac{\kappa}{4\pi}F_{i\,l}F^{\;l}_k+\left\{ \frac{
\kappa}{16\pi}F_{lm}F^{lm}+\Lambda-\kappa p-\right.$$
$$\left.-\frac{1}{2}\kappa g^{mn}\frac{
\partial\varphi}{\partial x^{\scriptscriptstyle m}}\frac{\partial\varphi}{
\partial x^{\scriptscriptstyle n}}\right\}g_{ik}.\eqno(11)
$$
Now we use suggestion (7) and as early consider massless scalar field
$\varphi$, so that
$$
\frac{\partial\varphi}{\partial x^{\scriptscriptstyle 0}}=\frac{\partial
\varphi}{\partial x^{\scriptscriptstyle 1}}=\frac{\partial\varphi}{\partial x
^{\scriptscriptstyle 2}}=0.
$$
Let only $F_{12}\neq0$. Then
$$
\frac{\kappa}{16\pi}F_{ lm}F^{lm}=\frac{\kappa}
{8\pi{\Omega}^{\lefteqn{\scriptstyle 2}}}\,
{(F_{12})}^2,
$$
$$
-\frac{\kappa}{4\pi}F_{1l}F^{\;l}_1=\frac{\kappa}{4\pi\Omega}\,{(F_{12})}^2,
$$
$$
-\frac{\kappa}{4\pi}F_{2l}F^{\;l}_2=\frac{\kappa}{4\pi\Omega}\,{(F_{12})}^2.
$$
By substituting these formulas and (2) into (11) we obtain
$$
\begin{array}{l}
{\displaystyle u_i=(\pm {\frac{1}{\mathstrut\sqrt{2\Omega}}},
\pm\sqrt{2\Omega}x
^{\scriptscriptstyle 2},\mp \sqrt{2\Omega}x^{\scriptscriptstyle 1},0)}\\
{\displaystyle  \frac{\mathstrut 3}{{\mathstrut}{\Omega}^{\lefteqn{
\scriptstyle2}}}=\frac{\kappa}{2\Omega}(c^2\rho+p)+\frac{1}{2\Omega}\left\{
\frac{\kappa}{8\pi{\Omega}^{\lefteqn{\scriptstyle2}}}\,{(F_{12})}^2+\Lambda-
\kappa p+\frac{1}{2}\kappa{\left(\frac{\partial\varphi}{\partial x^{
\scriptscriptstyle 3}}\right)}^2\right\}}\\
{\displaystyle \frac{{\mathstrut}6x^{\scriptscriptstyle 2}}
{\mathstrut\Omega}=
\kappa (c^2\rho+p)x^{\scriptscriptstyle 2}+x^{\scriptscriptstyle 2}\left\{
\frac{\kappa}{8\pi{\Omega}^{\lefteqn{\scriptstyle2}}}\,{(F_{12})}^2+\Lambda-
\kappa p+\frac{1}{2}\kappa{\left(\frac{\partial\varphi}{\partial x^{
\scriptscriptstyle 3}}\right)}^2\right\}}\\
{\displaystyle -\frac{{\mathstrut}6x^{\scriptscriptstyle 1}}
{\mathstrut\Omega
}=-\kappa (c^2\rho+p)x^{\scriptscriptstyle 1}-x^{\scriptscriptstyle 1}
\left\{
\frac{\kappa}{8\pi{\Omega}^{\lefteqn{\scriptstyle2}}}\,{(F_{12})}^2+\Lambda-
\kappa p+\frac{1}{2}\kappa{\left(\frac{\partial\varphi}{\partial x^{
\scriptscriptstyle 3}}\right)}^2\right\}}\\

{\displaystyle -{\mathstrut}12x^{\scriptscriptstyle 1}x^{\scriptscriptstyle 2
}=-2\kappa (c^2\rho+p)\Omega x^{\scriptscriptstyle 1}x^{\scriptscriptstyle 2}
-2\Omega x^{\scriptscriptstyle 1}x^{\scriptscriptstyle 2}\left\{\frac{
\mathstrut\kappa}{\mathstrut 8\pi{\Omega}^{\lefteqn{\scriptstyle2}}}\,
{(F_{12
})}^2+\Lambda-\kappa p+\frac{1}{2}\kappa{\left(\frac{\partial\varphi}
{\partial x^{
\scriptscriptstyle 3}}\right)}^2\right\}}\\
\end{array}
$$
$$
\begin{array}{l}
{\displaystyle {\mathstrut}12{x^{\scriptscriptstyle 2}}^2+2=
2\kappa\Omega (c^2
\rho+p){x^{\scriptscriptstyle 2}}^2+\frac{\mathstrut\kappa}{\mathstrut 4\pi
\Omega}\,{(F_{12})}^2+\Omega (2{x^{\scriptscriptstyle 2}}^2-1)\left\{\frac{
\kappa}{8\pi{\Omega}^{\lefteqn{\scriptstyle2}}}\,
{(F_{12})}^2+\Lambda-\right. }\\
          \hspace*{1.5cm}{\displaystyle \left.-\kappa
p+\frac{1}{2}\kappa{\left(\frac{\partial\varphi}
{\partial x^{\scriptscriptstyle 3}
}\right)}^2\right\}}\\
{\displaystyle {\mathstrut}12{x^{\scriptscriptstyle 1}}^2+2=
2\kappa\Omega (c^2
\rho+p){x^{\scriptscriptstyle 1}}^2+\frac{\kappa}{4\pi\Omega}\,{(F_{12})}^2+
\Omega (2{x^{\scriptscriptstyle 1}}^2-1)\left\{\frac{\mathstrut\kappa}{
\mathstrut 8\pi{\Omega}^{\lefteqn{\scriptstyle2}}}\,
{(F_{12})}^2+\Lambda-\right. }\\
       \hspace*{1.5cm}{\displaystyle \left.-
\kappa p+\frac{1}{2}\kappa{\left(\frac{\partial\varphi}{\partial x^{
\scriptscriptstyle 3}}\right)}^2\right\}}\\
\end{array}
$$
$$
\begin{array}{l}
{\displaystyle -\frac{\mathstrut 2}{\mathstrut\Omega}=
\kappa{\left(\frac{\partial
\varphi}{\partial x^{\scriptscriptstyle 3}}\right)}^2-
\left\{\frac{\kappa}{8\pi{
\Omega}^{\lefteqn{\scriptstyle2}}}\,{(F_{12})}^2+\Lambda-\kappa p+\frac{1}{2}
\kappa{\left(\frac{\partial\varphi}
{\partial x^{\scriptscriptstyle 3}}\right)}^2\right\}
}\\
{\displaystyle \frac{{\partial}^2\varphi}{\partial x^{\scriptscriptstyle 3}
\partial x^{\scriptscriptstyle 3}}=0}.\end{array}
$$
After nondifficult calculations the solution of the system can be written in
the form
$$
\left\{\begin{array}{l}{\displaystyle \Lambda=\kappa p}\\
{\displaystyle u_i=(\pm {\frac{1}{\sqrt{2\Omega}}},\pm\sqrt{2\Omega}x^{
\scriptscriptstyle 2},\mp\sqrt{2\Omega}x^{\scriptscriptstyle 1},0)}\\
{\displaystyle {(F_{12})}^2=
\frac{16\pi\Omega}{\kappa}+4\pi{\Omega}^2{\left(\frac{
\partial\varphi}{\mathstrut\partial x^{\scriptscriptstyle 3}}\right)}^2}\\
{\displaystyle \kappa c^2\rho=
\frac{4}{\Omega}-\Lambda-\kappa{\left(\frac{
\mathstrut\partial\varphi}{\mathstrut
\partial x^{\scriptscriptstyle 3}}\right)}^2}\\
{\displaystyle \frac{\mathstrut\partial\varphi}
{\mathstrut\partial x^{\scriptscriptstyle 3}}
=const, \frac{\mathstrut\partial\varphi}
{\mathstrut\partial x^{\scriptscriptstyle 0}}=
\frac{\mathstrut\partial\varphi}
{\mathstrut\partial x^{\scriptscriptstyle 1}}=
\frac{\mathstrut\partial\varphi}{\mathstrut\partial x^{\scriptscriptstyle 2
}}=0}\\
{\displaystyle 0\leq\Lambda\leq\frac{4}{\Omega}-
\kappa{\left(\frac{\mathstrut\partial
\varphi}{\partial x^{\scriptscriptstyle 3}}\right)}^2}.
\end{array}\right.\eqno (12)
$$

By using (10) we obtain for 4-vector of current
$$
j^{\,i}=(\frac{c}{4\pi\Omega}F_{12},0,0,0).
$$
Strength and induction of electric and magnetic fields are
\cite[p.331]{3}
$$
E_\alpha=0,\;D^\alpha=\left(\frac{2x^{\scriptscriptstyle 1}}
{\sqrt{2{\Omega}^{
\lefteqn{\scriptstyle 3}}}}F_{12},\frac{2x^{\scriptscriptstyle 2}}
{\sqrt{2{
\Omega}^{\lefteqn{\scriptstyle 3}}}}F_{12},0\right),
$$
$$
H_\alpha=\left(0,0,-\frac{1}{\sqrt{2{\Omega}^{\lefteqn{\scriptstyle 3}}}}\ \
F_{12}\right),\ \ B^\alpha=\left(0,0,-\frac{1}{\Omega}F_{12}\right).
$$

\section{Time machine}

The metric (1) admits the closed smooth timelike curves. 
For example we consider the following smooth closed curve
$$
L=
\{x^{\scriptscriptstyle 0}=const, x^{\scriptscriptstyle 1}=a\sin t,
x^{\scriptscriptstyle 2}=a\cos t,x^{\scriptscriptstyle 3}=const\}
$$
$$
a=const>\frac{1}{\sqrt{2}}.
$$
It is timelike in metric (1), that is $g_{ik}d{x}^id{x}^k>0$:
$$
g_{ik}d{x}^id{x}^k=a^2
\left\{g_{11}{\cos}^2t+g_{22}{\sin}^2t-2g_{12}\sin t\cos t\right\}=
a^2\Omega (2a^2-1)>0,
$$
as $a>1/\sqrt{2}$.

As  it known the distance which Time traveler must go, and his chronometric
invariant time are calculated with the help of following formulas \cite{3}:
$$
\tau(L)=\frac{1}{c}\oint\limits_{L}\frac{g_{0i}dx^{
\scriptscriptstyle i}}{\sqrt{g_{00}}}=\frac{2\pi a^2\sqrt{2\Omega}}{c},
$$
$$
l(L)=\oint
\limits_{L}\sqrt{\left(-g_{\alpha\beta} +\frac{g_{0\alpha}g_{0\beta}}
{g_{00}}\right)
\frac{dx^{\scriptscriptstyle\alpha}}{dt}
\frac{dx^{\scriptscriptstyle \beta}}{dt}}dt=2\pi
a\sqrt{\Omega}.\eqno (13)
$$
So "diameter"\ of domain which contains a Time machine $L$, has order
$l(L)\sim a\sqrt{\Omega}$. Proper time and time $\tau(l)$ are 
connected by  relation
$$
s(L)=\frac{1}{c}\oint\limits_{L}\sqrt{g_{ik}
\frac{dx^{\scriptscriptstyle i}}{dt}\frac{dx^{ \scriptscriptstyle k}}{dt}}dt=
\frac{1}{c}\oint\limits_{L}\sqrt{1-{\left(\frac{dl}{d\tau}\right)}^2}d\tau.
$$
The proper time goes to zero when
the velocity of  Time machine
goes to the velocity of light.

\section{Conclusion}

In the previous sections we showed that gravitational field, 
for which geometry of spacetime is discribed by metric (1), can be created 
by real phisical 
matter. But it is intresting how the our results and experimental data are
in agreement? 

At the begining we consider the cosmological solution. As it known in 
this case density of matter of the Universe is equal to 
$3\cdot 10^{-31}${\it g/cm}$^3$. If we wish no large distance for 
Time expedition  than $\tau(L)$ and $l(L)$ must be sufficiently small. 
By using (13) we conclude that
under considered assumption we must take small $\Omega$. In result, as our
solution discribe the Universe, using (8) and (12), we obtain
that scalar field which depend only on third coordinate must be
sufficiently large in both interpretations. (But only if in the first 
interpretation
the pressure of liquid is large). 

 And also we denote that in case of cosmological solution with small 
$\Omega$
in second interpritation the current and the strength and
induction of electric and magnetic fields are large.

If we take a solution with large density then considered in previous
paragraph scalar field $\theta=\theta(x^{\scriptscriptstyle 3})$ must be small.

Also we notice that in first interpretation we can remove the scalar field
$\theta$ or ideal liquid, and in second interpretation we can remove the
considering scalar field. The investigations of the results without analogous
scalar field were discribed in our  paper \cite{4}. 
The evaluations of the nessesary distance and chronometric invariant time for
Time travel in model without ideal liquid agree with the evaluations for
first interpretation of the stress-energy tensor in \cite{4}.
                        Moreover we denote that there exists 
interpretation of matter of gravitational field (1) as electric-magnetic
field and three real massless scalar field in ideal liquid. The
variants of matter which were considered by this article, are particular
cases of such interpretation.

\bigskip\bigskip
Author is grateful to A.K.Guts for recommended topic and consulting during
the work.

\end{document}